\documentstyle[12pt]{article}
\begin{document}
\pagestyle{empty}
\begin{flushright}
   {CERN-TH.6411/92} \\
\end{flushright}
\vspace* {10mm}
\renewcommand{\thefootnote}{\fnsymbol{footnote}}
\begin{center}
   {\bf THE POKROVSKI-TALAPOV PHASE TRANSITION \\
        AND QUANTUM GROUPS
   \footnote{To appear in the Proceedings of
   the II International Wigner Symposium,
   Goslar 1991}} \\[13MM]
   Haye Hinrichsen \\[5mm]
   {\it Universit\"{a}t Bonn,
   Physikalisches Institut \\
   Nussallee 12, W-5300 Bonn 1, FRG} \\[10mm]
   Vladimir Rittenberg \\[5mm]
   {\it Theory Division, CERN \\
   CH-1211 Geneva 23, Switzerland} \\[4cm]
	  {\bf Abstract}
\end{center}
\renewcommand{\thefootnote}{\arabic{footnote}}
\addtocounter{footnote}{-1}
\vspace*{2mm}
We show that the XY quantum chain
in a magnetic field is invariant under a two
parameter deformation of the $SU(1/1)$
superalgebra. One is led to an extension
of the braid group and the Hecke
algebras which reduce to
the known ones when the
two parameter coincide. The
physical significance of the two
parameters is discussed. When both are equal to one, one
gets a Pokrovski-Talapov phase transition. We also show
that the representation theory of the quantum
superalgebras indicates how to take the appropriate
thermodynamical limits.
\vspace{3cm}  \begin{flushleft}
   CERN-TH.6411/92 \\
   February 1992 \\
\end{flushleft}
\thispagestyle{empty}
\mbox{}

\newpage
\setcounter{page}{1}
\pagestyle{plain}

\section{Introduction}

There were several attempts to extend the one-parameter
quantum algebras to multiparameter ones \cite{r1}. As shown
however by Reshetikhin \cite{r2} the link polynomials
depend only on one parameter. One can state this result in
a different way: if one has a one-dimensional quantum chain
which is invariant under a multiparameter quantum algebra,
one can do a similarity transformation which eliminates all
the parameters but one. As will be shown in this paper, the
situation is different in the case of quantum
superalgebras. We will start with a physical example.
Consider the quantum chain

\begin{equation}
H\;=\;\Delta_{q}\sum_{i=1}^{L}\sigma_{i}^{z}\;
+\;\frac{\Delta_{\eta}}{2}\sum_{i=1}^{L-1}
\:[(1+u)\sigma_{i}^{x}\sigma_{i+1}^{x}\:
+\: (1-u)\sigma_{i}^{y}\sigma_{i+1}^{y}]\;+\;B\;+\;S,
\label {e1}
\end{equation}

\noindent
where $\sigma^{x}$, $\sigma^{y}$
and $\sigma^{z}$ are Pauli matrices
inserted in the i-th position
of the Kronecker product

\begin{eqnarray}
& & \sigma^{k}_{i}\;=\;\bf{1}\otimes\bf{1}
\otimes\ldots\otimes
\underbrace{\sigma^{k}}_{i} \otimes
\ldots\otimes\bf{1}\otimes\bf{1}
\;\;\;\;\;\;\;\;\;\;\;\;\;\;
\rm{(i=1,2,\ldots L)}
\label{e2}
\\ \,
& & [\sigma_{i}^{k},\sigma_{j}^{l}]\; = \;0.
\;\;\;\;\;\;\;\;\;\;\;\;\;\;
\;\;\;\;\;\;\;\;\;\;\;\;\;\;\;\;\;\;\;\;\;\;\;\;\;
\;\;\;\;\;\;\;\;\;\;\;\;\;\;
 \rm{(i\neq j)}\nonumber
\end{eqnarray}

\noindent
$\Delta_{q}$, $\Delta_{\eta}$
and $u$ are parameters, $B$ and $S$ are boundary
and surface terms respectively.
This chain appears in the domain wall theory
of two-dimensional commensurate-incommensurate
phase transitions \cite{r3,r10}
and in Glauber's kinetic Ising model \cite{r4}.
In order to make contact with quantum algebras
we will first make an important
change of notations, choose $B=0$
(no periodic boundary conditions!)
and fix $S$ by

\begin{eqnarray}
\Delta_{q}\;=\;\frac{q+q^{-1}}{2},\;\;\;\;\;\;
& & \Delta_{\eta}\;=\;\frac{\eta+\eta^{-1}}{2},
\;\;\;\;\;\;\;\;\;\;\;
u\;=\;\frac{\eta-\eta^{-1}}{\eta+\eta^{-1}}
\label{e3} \\ \, \nonumber \\
& S & \;=\;\frac{1}{2}\:
(q^{-1}\sigma_{1}^{z}+q\:\sigma_{L}^{z}).\nonumber
\end{eqnarray}

\noindent
With this change of notations we have

\begin{eqnarray}
& & H \; = \; H(q,\eta) \; = \;
\sum_{i=1}^{L-1}\:H_{i}(q,\eta)
\label {e4}
\\
& & H_{i}(q,\eta) \; = \;
\frac{1}{2}\: [\eta\:\sigma_{i}^{x}\sigma_{i+1}^{x}
\; + \; \eta^{-1}\sigma_{i}^{y}\sigma_{i+1}^{y}
\; - \; q\,\sigma_{i}^{z} \; -
\; q^{-1}\sigma_{i+1}^{z} ].\nonumber
\end{eqnarray}

\noindent
A detailed discussion of the properties
of the chain given by eq. (\ref{e4})
will be given elsewhere~\cite{r5},
here we are going to mention only a few.
First, there are the symmetry properties

\begin{equation}
\label{e5}
H(q,\eta) \; \doteq \; H(q^{-1},\eta)
\; \doteq \; H(q,\eta^{-1})
\; \doteq \; H(\eta,q).
\end{equation}

\noindent
The ''equality'' among the Hamiltonians
implies that the spectra are identical.
The first two equalities are obvious
but not the last one which reminds of
duality transformations of
quantum chains \cite{r6}. In the continuum limit,
one has the following phase structure \cite{r3,r5}:

\begin{tabbing}
\ \ \ \ \ \ \ \ \ \= $\Delta_{q}
\leq 1$,\ \ \  $\Delta_{\eta}\leq 1$:\ \ \
\ \ \ \ \ \ \ \ \ \ \ \ \ \ \
\ \ \ \ \ \ \ \  \ \ \ \ \
\=massless-incommensurate
\\

\>$\Delta_{q}\leq 1$,\ \ \
$\Delta_{\eta}>1$\ or \ $\Delta_{q}>1$,\ \ \
$\Delta_{\eta}\leq 1$:\ \ \
\> massive incommensurate \\

\>$\Delta_{q}>1$,\ \ \
$\Delta_{\eta}>1$,\ \ \
$\Delta_{q} \neq \Delta_{\eta}$: \
\ \ \> massive\\

\>$\Delta_{q}=\Delta_{\eta}$,
\ \ \ $(\Delta_{q} > 1)$ : \
\ \ \>critical Ising type \\

\>$\Delta_{q}=\Delta_{\eta}=1$ :
\ \ \>Pokrovsky-Talapov phase transition \cite{r11} \\
\end{tabbing}

\noindent
We will return to the physical picture later on
in the text.
It is by now clear that the
properties of the chain depend on \underline{both}
parameters $q$ and $\eta$. \\
\indent
We now perform a Jordan-Wigner transformation.
First write $\sigma_{j}^{z}\;=\;
-i\sigma_{j}^{x}\sigma_{j}^{y}$ and next define

\begin{eqnarray}
\label{e6}
& & \tau_{j}^{x,y}\;=\;\exp(
{\frac{i \pi}{2} \sum_{k=1}^{j-1}
(\sigma_{k}^{z}+1)})
\;\sigma_{j}^{x,y} \\
& & \{\tau_{i}^{x},\tau_{j}^{x}\} \;=\;
    \{\tau_{i}^{y},\tau_{j}^{y}\} \;=\;
    \{\tau_{i}^{x},\tau_{j}^{y}\} \;=\; 0.
\ \ \ \ \ \ \ \ \ \ \ \ \ \ \ \ \ \ (i \neq j) \nonumber
\end{eqnarray}

\noindent
Using Eq. (\ref{e4}) and (\ref{e6}) we get

\begin{equation}
H_{j} \;=\; \frac{i}{2}
[\:\eta\tau_{j}^{y}\tau_{j+1}^{x}\: -
\:\eta^{-1}\tau_{j}^{x}\tau_{j+1}^{y} \: +
\: q \tau_{j}^{x}\tau_{j}^{y} \: +
\: q^{-1}\tau_{j+1}^{x}\tau_{j+1}^{y}].
\end{equation}

\noindent
We now observe the following important identity

\begin{equation}
\label{e7}
[T^{X},H(q,\eta)] \;=\; [T^{Y},H(q,\eta)] \;=\; 0
\end{equation}

\noindent
with

\begin{eqnarray}
\label{v9}
& & T^{X} \;=\; \Delta(\tau^{x}) \;=\;
    \alpha^{-\frac{L+1}{2}} \;\sum_{j=1}^{L}
    \alpha^{j} \tau_{j}^{x}
\nonumber \\
& & T^{Y} \;=\; \Delta(\tau^{y})\;=\;
    \beta^{-\frac{L+1}{2}} \;\sum_{j=1}^{L}
    \beta^{j} \tau_{j}^{y}
\end{eqnarray}

\begin{equation}
\label{v10}
    \{T^{X},T^{X}\} \;=\; 2[L]_{\alpha}, \ \ \ \ \ \
    \{T^{Y},T^{Y}\} \;=\; 2[L]_{\beta}, \ \ \ \ \ \
    \{T^{X},T^{Y}\} \;=\; 0,
\end{equation}

\noindent
where $L$ is the length of the chain and

\begin{equation}
\label{e9}
\alpha \;=\; \frac{q}{\eta}, \ \ \ \ \ \ \ \
\beta  \;=\; q \eta, \ \ \ \ \ \ \ \
[L]_{\lambda} \;=\;
\frac{\lambda^{L}-\lambda^{-L}}{\lambda-\lambda^{-1}}.
\end{equation}

\noindent
The equalities (\ref{e7}) come from
the existence of a fermionic zero mode for any
$q$ and $\eta$. The equations (\ref{v10})
together with the coproduct (\ref{v9})
give a representation of a Hopf algebra.
Before we proove this statement let us
consider the case $\alpha=\beta=q$.

\section{The $\eta=1$ case. Mathematics.}
We first notice that in this case
$S^{z}=\frac{1}{2}\:\sum_{i=1}^{L}
\sigma_{i}^{z}$ also commutes with $H(q,\eta)$.
We now remind the reader the
$U_{\alpha}[SU(1/1)]$ algebra \cite{r7}.
With $A^{\pm}= \frac{1}{2}(T^{X}\pm iT^{Y})$ we have

\begin{eqnarray}
\label{e10}
& & \{A^{\pm},A^{\pm}\}=0,
\ \ \ \ \ \ \{A^{+},A^{-}\}=[E]_{\alpha}, \ \ \ \ \
\ \,[S^{z},A^{\pm}] = \pm A^{\pm} \\
& & [E,S^{z}]=[E,A^{\pm}]=0 \nonumber
\end{eqnarray}

\noindent
with the coproduct

\begin{eqnarray}
\label{e11}
& & \Delta(\alpha,A^{\pm}) \;=\;
\alpha^{E/2} \otimes A^{\pm} +
       A^{\pm}\otimes\alpha^{-E/2}  \nonumber \\
& & \Delta(\alpha,S^{z}) \;\;=\;
S^{z}\otimes {\bf1} + {\bf1}\otimes S^{z} \\
& & \Delta(\alpha,E) \;\;\;=\; E
\otimes {\bf1} \; + {\bf1} \otimes E. \nonumber
\end{eqnarray}

\noindent
The fermionic representations
correspond to take $E={\bf1}$, \
$S^{z}=\frac{1}{2}\sigma^{z}$,
\ $A^{\pm}=a^{\pm}$ and $\{a^{+},a^{-}\}=1$ in Eq
(\ref{e11}). In this representation
$E$ in Eq. (\ref{e10}) is equal to $L$ (the
number of~sites). Comparing now
(\ref{e10}), (\ref{e11}) with
Eqs. (\ref{v9},\ref{v10}) we
observe \cite{r8} that the
quantum chain (\ref{e4}) with $\eta=1$ is invariant
under $U_{\alpha}[SU(1/1)]$
transformations. It was
also shown by Saleur \cite{r8}
that the quantities $U_{j}=\Delta_{q}-H_{j}(q,1)$
are the generators of the Hecke
algebra

\begin{eqnarray}
\label{e12}
& & U_{j}^{2} \;=\; 2\:\Delta_{q}\:U_{j} \nonumber \\
& & U_{j}U_{j\pm 1}U_{j} -
U_{j} \;=\; U_{j\pm 1}U_{j}U_{j \pm 1} - U_{j \pm 1} \\
& & U_{i}U_{i\pm j} \;=\;
U_{i \pm j}U_{i}. \ \ \ \ \ \ \ \ \ \ \ \ \ \
\ \ \ \ \ \ \ \ \ \ \ \ \ \
\ \ \ \ \ \ \ \ \ \ \ \ \ \ \ \ \ (j \neq 1) \nonumber
\end{eqnarray}

\noindent
Actually they correspond to
a quotient of this algebra since the generators
satisfy also the relations \cite{r9}

\begin{equation}
\label{e13}
U_{j}U_{j+2}U_{j+1}(2\Delta_{q}-U_{j})
(2 \Delta_{q} -U_{j+2}) \;=\;0.
\end{equation}

\noindent
The generators $\check{R}_{j} =
\frac{q-q^{-1}}{2} + H_{j}(q,1)$ satisfy the
braiding relations

\begin{equation}
\label{e14}
\check{R}_{j} \check{R}_{j\pm 1}
\check{R}_{j} \;=\;
\check{R}_{j\pm 1} \check{R}_{j}
\check{R}_{j\pm 1}
\end{equation}

\noindent
with

\begin{equation}
\label{e15}
\check{R}_{j}^{2} \;=\;
(q-q^{-1})\:\check{R}_{j}\:+\: 1.
\end{equation}

\noindent
Considering the matrices $R_{j}=P
\check{R}_{j}$ ($P$ is the graded
permutation operator) we have \\
(see~Eq.~(\ref{e11}))

\begin{equation}
\label{e16}
R \Delta(\alpha)R^{-1} \;=\; \Delta(\alpha^{-1}).
\end{equation}

\section{The $ \eta = 1 $ case. Physics.}
Before persuing our mathematical developements let us
pause and discuss some physical implications. First we
notice the very unusual role of the operator $E$ for the
quantum chain. It does not behave like an usual symmetry
operator in quantum mechanics (like the angular momentum)
which commutes with the Hamiltonian and helps in its
diagonalisation. Since $E$ simply counts the number of
sites it plays a different role that we clarify now. From
Eq. (\ref{e10}) we see that for $\alpha$ generic
($\alpha\neq e^{i\pi\frac{r}{s}}$), $U_\alpha[SU(1/1)]$ has
two-dimensional irreducible representations and one
one-dimensional irreducible representation where
$A^{\pm}=E=S^z=0$. If $\alpha$ is not generic
($\alpha=e^{i\pi\frac{r}{s}}$), notice that

\begin{equation}
\label{con1}
\{ A^+,A^- \} =
\frac{\sin(\frac{\pi rL}{s})}{\sin(\frac{\pi r}{s})}
\end{equation}

\noindent
and that for $L=ns$ one has only one-dimensional
irreducible representations. This implies that for a given
value of $q$, changing $L$ one can reach pathological
situations. As shown in Ref.~\cite{r5}, if $L=ns$ one has
not only one zero mode but two which makes the
degeneracies larger and not smaller as one would expect from
the fact that we have only one-dimensional irreducible
representations. In order to avoid this type of problems
and to keep the normalisations of the zero-mode operator
(i.e. $A$ and $A^+$), if one wants to take the
thermodynamical limit, one has to take sequences like

\begin{equation}
\label{con2}
L = ns+t \;\;\;\;\;\;\;\;\;\;\;\;
(t=0,1, \ldots, s-1;\; n \in Z_+)
\end{equation}

\noindent
and the results will depend on the sequence. The
necessity of taking sequences for the quantum chain
(\ref{e1}) with periodic boundary conditions is already
known \cite{r12} but now we understand its origin. The
same observation applies when we have two parameters (see
Eq. (\ref{v9})) and one or both of them are not generic
\cite{r5}. \\
\indent
A more detailed discussion of the physical meaning of the
parameter $q$ as well as the connection of the model with
the experimental data \cite{r12x} can be found in Ref.
\cite{r5}.

\section{The $\eta \neq 1$ case.}
 \noindent
As suggested by Eqs.
(\ref{v9},\ref{v10})
we define the two parameter deformation of the
$SU(1/1)$ algebra as follows:

\begin{eqnarray}
\label{e17}
 & \{T^{X},T^{X}\} \;=\; 2\:[E]_{\alpha},
\ \ \ \ \ \ \ \
 & \{T^{Y},T^{Y}\} \;=\; 2\:[E]_{\beta} \\
 & \{T^{X},T^{Y}\} \;=\; 0
\ \ \ \ \ \ \ \ \ \ \ \ \ \ \
 & [E,T^{X}] \;=\; [E,T^{Y}] \;=\; 0 \nonumber
\end{eqnarray}

\noindent
with the coproduct

\begin{eqnarray}
\label{e18}
& & \Delta(\alpha,\beta;T^{X}) \;=\;
\alpha^{E/2}\otimes T^{X} \;+\;
T^{X}\otimes\alpha^{-E/2} \nonumber \\
& & \Delta(\alpha,\beta;T^{Y}) \;=\;
\beta^{E/2}  \otimes T^{Y} \;+\;
T^{Y}\otimes\beta^{-E/2} \\
& & \Delta(\alpha,\beta;E) \;=\; E
\otimes {\bf1} \;+\; {\bf1} \otimes E. \nonumber
\end{eqnarray}

\noindent
Notice that $S^{z}$ does not appear
in the algebra anymore. We denote this
quantum algebra by $U_{\alpha,\beta}
[SU(1/1)]$. It is a Hopf algebra for the same
reasons as the $U_{\alpha}[SU(1/1)]$.
If we take the fermionic
representations $E=1$, \ $\tau^{x}=(a^{+}+a^{-})$, \
\ $\tau^{y}=-i(a^{+}-a^{-})$, from Eq.
(\ref{e18}) we derive
Eqs. (\ref{v9},\ref{v10}). The
quantum chain $H(q,\eta)$ is
thus invariant under the quantum algebra
$U_{\alpha,\beta}[SU(1/1)]$.
We would like to see what replaces the relations
(\ref{e12}-\ref{e16}) when
we have two parameters. We first notice a remarkable
identity satisfied by the $H_{j}(q,\eta)$

\begin{eqnarray}
\label{e19}
& & \left[ H_{j}H_{j\pm1}H_{j}-
H_{j\pm1}H_{j}H_{j\pm1} + (\nu-1)
(H_{j}-H_{j\pm1}) \right] \:
(H_{j}-H_{j\pm1}) = \mu \\ \nonumber \\
& & \ \ \ \ \ \ \ \ \ \ \ \ \ \ \ \ \ \ \ \  \ \
\ \ \ \ \ \ \ \ \ \ \ \ \
H_{j}^{2} \;=\; \nu, \nonumber
\end{eqnarray}

\noindent
where

\begin{eqnarray}
\label{e20}
& & \nu \;=\; \left(\frac{\alpha+\alpha^{-1}}{2}\right)
    \left(\frac {\beta+\beta^{-1}}{2}\right)
\;=\; {\left(\frac{q+q^{-1}}{2}\right)}^{2} +
    {\left(\frac{\eta+\eta^{-1}}{2}\right)}^{2} -1 \\
& & \mu \;=\;
    {\left( \frac{\alpha+\alpha^{-1}}{2}-
       \frac{\beta+\beta^{-1}}{2} \right)}^{2} \;=\;
    4\:{\left(\frac{q-q^{-1}}{2}\right)}^{2}\;
    {\left(\frac{\eta-\eta^{-1}}{2}\right)}^{2}.
\nonumber  \end{eqnarray}

\noindent
We can now define a generalised Hecke algebra taking

\begin{eqnarray}
\label{e21}
& & U_{i} \;=\; \sqrt{\nu}-H_{i}(p,q) \nonumber \\
& & (U_{i}U_{i\pm1}U_{i}-
U_{i\pm1}U_{i}U_{i\pm1}-U_{i}+U_{i\pm1}) \;
(U_{i}-U_{i\pm1}) \;=\; \mu \\
& & U_{i}^{2} \;=\; 2\:\sqrt{\nu}\:U_{i}. \nonumber
\end{eqnarray}

\noindent
Notice that when $\eta=1$, \
$\mu=0$ and we have representations which coincide
with those of
the original Hecke algebra
(For a detailed discussion of the representation
theory of (\ref{e21}) see Ref.~\cite{r15}).
We did not have the patience
to find the equivalent of Eq.~(\ref{e13}) which
gives the quotient of the
generalised Hecke algebra
(\ref{e21}) corresponding to
the chain given by Eq. (\ref{e4}).
Another quotient is however suggested by the
structure of Eq. (\ref{e21}):

\begin{equation}
\label{e22}
(U_{i}U_{i\pm1}U_{i}-U_{i})\:
(U_{i}-U_{i\pm1}) \;=\; \frac{\mu}{2}.
\end{equation}

\noindent
For $\mu=0$ one gets in this case representations of
the Temperley-Lieb algebra $U_iU_{\pm i}U_i = U_i$.
We now turn our
attention to the generalised
braid group algebra. We take
$\check{R}_{i} = H_{i}(q,\eta)+\sqrt{\nu-1}$ and get

\begin{equation}
\label{e23}
(\check{R}_{i}\check{R}_{i\pm1}\check{R}_{i} -
 \check{R}_{i\pm1}\check{R}_{i}\check{R}_{i\pm1}) \:
(\check{R}_{i}-\check{R}_{i\pm1}) \;=\; \mu
\end{equation}

\noindent
with

\begin{equation}
\label{e24}
\check{R}_{i}^{2} \;=\; 1+\sqrt{\nu-1}\:\check{R}_{i}\;.
\end{equation}

\noindent
In the basis where the $\sigma_{i}^{z}$
are diagonal (see Eq. (\ref{e4}))
we have

\begin{equation}
\label{e25}
\check{R}_{i} \;=\; \left(
\begin{array}{cccc}
\sqrt{\nu-1}+\frac{q+q^{-1}}{2}
&  0  &  0  &  \frac{\eta-\eta^{-1}}{2} \\
0 & \sqrt{\nu-1}-\frac{q-q^{-1}}{2}
& \frac{\eta+\eta^{-1}}{2} & 0 \\
0 & \frac{\eta+\eta^{-1}}{2} &
\sqrt{\nu-1}+\frac{q+q^{-1}}{2} & 0 \\
\frac{\eta-\eta^{-1}}{2} & 0 & 0
& \sqrt{\nu-1}-\frac{q+q^{-1}}{2}
\end{array}
\right).
\end{equation}

\noindent
We take the graded permutation matrix P

\begin{equation}
\label{e26}
P \;=\; \left(
\begin{array}{cccc}
\;1 & & & \\
& \;0 & \;1 & \\
& \;1 & \;0 & \\
& & & -1
\end{array}
\right)
\end{equation}

\noindent
and define the matrix $R_{i}=P\check{R}_{i}$.
We now write the coproduct
(\ref{e18}) in the original language of Pauli matrices

\begin{eqnarray}
\label{e27}
& & \Delta(\alpha,\beta;T^{X}) \;=\;
\alpha^{-1/2}(\sigma^{y}\otimes\bf{1}) \;-\;
\rm{\alpha^{1/2}
(\sigma^{z}\otimes\sigma^{y})} \nonumber \\
& & \Delta(\alpha,\beta;T^{Y}) \;=\;
\beta^{-1/2}(\sigma^{x}\otimes\bf{1}) \;-\;
\rm{\beta^{1/2} (\sigma^{z}\otimes\sigma^{x})} \\
& & \Delta(\alpha,\beta;E)
\;=\; \bf{1}\otimes\bf{1} \;+\; \bf{1}\otimes\bf{1}.
\nonumber
\end{eqnarray}

\noindent
It is trivial to check that similar
to Eq. (\ref{e16}) we get

\begin{equation}
\label{e28}
R\:\Delta(\alpha,\beta)\:R^{-1}
\;=\; \Delta(\alpha^{-1},\beta^{-1}).
\end{equation}

\section{Are more-parameter deformations possible?}
\noindent
In this section we would like
to show that for $U_{\alpha,\beta}[SU(1/1)]$
one can introduce more than
two parameters (as in the Lie algebra case when we
had more than one). The most
general chain which has a zero mode for all its
values of the parameters is \cite{r5}

\begin{eqnarray}
\label{e29}
H_{i}\;\;=\; & \;\frac{1}{2} \: \{ \: &
\frac{\Theta+\Theta^{-1}}{2}
(\eta\zeta\sigma_{i}^{x}\sigma_{i+1}^{x}
+ \eta^{-1}\zeta^{-1}
\sigma_{i}^{y}\sigma_{i+1}^{y}) \nonumber \\
& \;\;+ & \frac{\Theta-\Theta^{-1}}{2}
(\eta\zeta^{-1}\sigma_{i}^{x}\sigma_{i+1}^{y} +
\eta^{-1}\zeta\sigma_{i}^{y}\sigma_{i+1}^{x}) \\
& \;\;+ & q\:\sigma_{i}^{z} \;+\;
q^{-1}\sigma_{i+1}^{z}) \: \}. \nonumber
\end{eqnarray}

\noindent
$H_{i}$ depends on four parameters.
One can check however that Eq. (\ref{e19})
holds with

\begin{eqnarray}
\label{e30}
\nu \; & = & \; \frac{q^{2}+q^{-2}}{4} +
				\frac{(\eta^{2}-\eta^{-2})
    (\zeta^{2}-\zeta^{-2})}{8} +
    \frac{(\eta^{2}+\eta^{-2})
    (\zeta^{2}+\zeta^{-2})(\Theta^{2}+\Theta^{-2})}{16},
    \nonumber \\
    \nonumber \\
\mu \; & = & \;
    \left(
    \frac{(\eta^{2}-\eta^{-2})
    (\zeta^{2}-\zeta^{-2})}{4} +
    \frac{(\eta^{2}+\eta^{-2})
    (\zeta^{2}+\zeta^{-2})
    (\Theta^{2}+\Theta^{-2})}{8} - 1
    \right)
    \\
& & \ \times \ \
    \left(
    \frac{{(q-q^{-1})}^{2}}{2} +
    \frac{{(\Theta-\Theta^{-1})}^{2}
    {(\frac{\eta}{\zeta}-\frac{\zeta}{\eta})}^{2}}{8}
    \right)
    \nonumber
\end{eqnarray}

\noindent
which means that we are back to two parameters.
This means that there is a similarity transformation which
connects the Hamiltonian with four parameters and the one
with two (see Eq.~(\ref{e4})). In order to illustrate this
point we consider the ''two-parameter deformation'' of Ref.
\cite{r13}. It corresponds to the choice

\begin{equation}
\label{con3}
\zeta=e^{i\pi/4}, \;\;\;\;\;\;\;
\eta=e^{-i\pi/4}, \;\;\;\;\;\;\;
q=\sqrt{QP}, \;\;\;\;\;\;
\theta = \sqrt{\frac{Q}{P}}
\end{equation}

\noindent
in Eq. (\ref{e29}) where $Q$ and $P$ are the two parameters given in
\cite{r13}. From Eq. (\ref{e30}) we get $\mu=0$ which implies that we
are back to the $U_\alpha[SU(1/1)]$~case. From Eq. (\ref{e29}) we
derive

\begin{eqnarray}
\label{con4}
\check{R}_i & = &
\frac{1}{2} (\sqrt{QP}-\frac{1}{\sqrt{QP}}) +
\sqrt{\frac{Q}{P}}\,\sigma_i^+\sigma_{i+1}^- +
\sqrt{\frac{P}{Q}}\,\sigma_i^-\sigma_{i+1}^+ + \\
& & \frac{1}{2} \sqrt{QP}\, \sigma_i^z +
\frac{1}{2} \frac{1}{\sqrt{QP}}\, \sigma_{i+1}^z
\; . \nonumber
\end{eqnarray}

\noindent
We now do the similarity transformation \cite{r14}

\begin{equation}
\label{con5}
\sigma_i^+ \rightarrow (\sqrt{Q}{P})^{i-1}\,\sigma_i^+,
\;\;\;\;\;\;
\sigma_i^- \rightarrow (\sqrt{P}{Q})^{i-1}\,\sigma_i^-,
\;\;\;\;\;\;
\sigma_i^z \rightarrow \sigma_i^z
\end{equation}

\noindent
and recover Eq. (\ref{e4}) with $\eta=1$ and
$q=\sqrt{QP}$, which means that the two-parameter deformation is a
one-parameter deformation.

\end{document}